\def\etal{{\rm et al.}}

\newcommand{\be}{\begin{equation}}
\newcommand{\ee}{\end{equation}}
\newcommand{\musun}{\mu_{\odot}}
\newcommand{\muearth}{\mu_{\oplus}}
\documentclass[namedreferences]{kluwer}

\usepackage{graphicx}
\setkeys{Gin}{width=1.05\textwidth}

\begin{document}
\begin{article}

\begin{opening}

\title{Astronomical engineering:\\
a strategy for modifying planetary orbits}

\author{D. G. \surname{Korycansky}\email{kory@earthsci.ucsc.edu}}
\institute{CODEP, Department of Earth Sciences, University of California,
Santa Cruz, CA 95064 USA}
\author{Gregory \surname{Laughlin}\email{gpl@acetylene.arc.nasa.gov}}
\institute{NASA Ames Research Center, 245-3 Moffett Field, CA 94035 USA}
\author{Fred C. \surname{Adams}\email{fca@umich.edu}}
\institute{Physics Department, University of Michigan, Ann Arbor, MI 48109 USA}

\runningtitle{Astronomical Engineering}
\runningauthor{Korycansky \etal}


\begin{ao}\\
D. G. Korycansky (kory@earthsci.ucsc.edu)\\
CODEP, Department of Earth Sciences\\
University of California\\
Santa Cruz, CA 95064 USA\\
\end{ao}

\begin{abstract} 

The Sun's gradual brightening will seriously compromise the Earth's
biosphere within $\sim10^9$ years.  If Earth's orbit migrates outward,
however, the biosphere could remain intact over the entire
main-sequence lifetime of the Sun. In this paper, we explore the
feasibility of engineering such a migration over a long time
period. The basic mechanism uses gravitational assists to (in effect)
transfer orbital energy from Jupiter to the Earth, and thereby
enlarges the orbital radius of Earth.  This transfer is accomplished
by a suitable intermediate body, either a Kuiper Belt object or a main
belt asteroid. The object first encounters Earth during an inward pass
on its initial highly elliptical orbit of large ($\sim 300$ AU)
semimajor axis. The encounter transfers energy from the object to the
Earth in standard gravity-assist fashion by passing close to the
leading limb of the planet. The resulting outbound trajectory of the
object must cross the orbit of Jupiter; with proper timing, the
outbound object encounters Jupiter and picks up the energy it lost to
Earth. With small corrections to the trajectory, or additional
planetary encounters (e.g., with Saturn), the object can repeat this
process over many encounters. To maintain its present flux of solar
energy, the Earth must experience roughly one encounter every 6000
years (for an object mass of $10^{22}$ g). We develop the details of
this scheme and discuss its ramifications.  
\end{abstract} 

\end{opening}

\section{Introduction}		\label{sec:intro}

As the Sun burns through its hydrogen on the main sequence, it
steadily grows hotter, larger, and more luminous. Stellar evolution
calculations show that in $\sim$1.1 billion years the Sun will be 11\%
brighter than it is today (e.g., \opencite{SBK}).
Global climate models indicate that such an increase in
insolation would drive a ``moist greenhouse'' on the Earth (\opencite{K};
\opencite{NHA}) which will have a catastrophic
effect on the surface biosphere. In 3.5 billion years, the total
luminosity of the Sun will be 40\% larger than the present
value. Under such conditions, the Earth will undergo a catastrophic
``runaway greenhouse'' effect \cite{K}, which will likely spell a
definite end to life on our planet.

Although the Earth's ecosystem will be seriously compromised within a
billion years, the Sun is presently less than halfway through its main
sequence life.  Indeed, in 6.3 billion years, the luminosity of the
Sun is expected to be ``only'' a factor of 2.2 greater than its
current value. At that time, a planet located at 1.5 AU from the Sun
would receive the same flux of solar energy that is now intercepted by
the Earth.

If the radius of the Earth's orbit were somehow to be gradually
increased, catastrophic global warming could be avoided, and the
lifespan of the surface biosphere could be extended by up to five
billion years. In this paper, we study the feasibility of altering
planetary orbits over long time scales.  Special attention will be
given to the specific case of the Earth, but many of the issues we
address are of more general astronomical and astrobiological
interest.

The present orbital energy of the Earth is $-2.7 \times 10^{40}$ erg.
Moving the Earth to a circular orbit of 1.5 AU radius would require
$8.7 \times 10^{39}$ erg.  An attractive scenario for gradually
increasing the Earth's orbital radius is to successively deflect a
large object or objects from the outer regions of the solar system
(the Oort Cloud or the Kuiper Belt) onto trajectories which pass close
to the Earth. By analogy to the gravity-assisted flight paths employed
by spacecraft directed to outer solar-system targets (e.g., \opencite{BA2},
\opencite{M}), the close passage of such an object to
the Earth can result in a decrease in the orbital energy of the object
and a concomitant increase of the Earth's orbital energy.  For optimal
trajectories which nearly graze the Earth's atmosphere, the energy
boost imparted to the Earth is $2.4 \times 10^{12}$ erg gm$^{-1}$ of
object mass \cite{N}.  Work by \inlinecite{ST}
suggests that even bodies that are weakly held together (``rubble
piles'') can survive passages that approach less than 1 Earth radius
from the Earth's surface, allowing energy transfers of $\sim 10^{12}$
erg gm$^{-1}$.

Typical masses for large Kuiper belt objects are of the order of
$10^{22}$ grams, meaning that roughly $10^{6}$ passages (involving a
cumulative flyby mass of approximately 1.5 Earth masses) would be
needed to move the Earth out to 1.5 AU. Thus, over the remaining
lifespan of the Sun, approximately one passage every 6000 years on
the average would be required.

The outer reaches of the Solar System contain an ideal reservoir of
material which could be used to move the Earth. The Kuiper Belt is
populated by a large number of objects that are larger than 100 km in
diameter; the Kuiper belt may contain as many as 10$^{5}$ such bodies,
totaling perhaps 10\% of the Earth's mass \cite{J}, although
these numbers remain uncertain.  The Oort cloud is believed to contain
about $10^{11}$ objects totaling 30 or more Earth masses (see, e.g.,
\opencite{W}).  As evidenced by the frequent passage of long period
Sun-grazing comets originating in this region, many Oort cloud objects
would need only small trajectory changes in order to bring them
into appropriate Earth-crossing orbits. Indeed, strategies for
modifying the orbits of asteroids and comets have been extensively
discussed in the context of mitigating the hazard posed by such
objects impacting the Earth (see, e.g., \opencite{AH}, \opencite{MNZ},
\opencite{S}).
Alternatively, a main belt object could be deflected into
an orbit which has an aphelion in the outer solar system.

Our approach in this paper is as follows: In \S 2, we discuss the
details of our gravity assist scheme. This scheme uses an asteroid or
large comet as a catalyst to transfer orbital angular momentum and
energy from Jupiter to the Earth. We investigate the energy
requirements of the scheme, the nature of the course corrections
demanded, and also the needed accuracy. In \S 3, we discuss
additional considerations, such as long term orbital stability,
complications produced by other planets, and larger issues.  We
present our conclusions in \S 4. Although this problem raises many
possible interesting (and rather speculative) issues, the present
paper discusses only a few of them.

\section{The gravity-assist scheme} 
\label{sec:assist}

As mentioned in the introduction, our underlying scenario uses
repeated gravity assists to (in effect) transfer orbital energy from
Jupiter to the Earth, thus enlarging the Earth's orbit and reducing
the received solar flux. Multiplanet encounter trajectories have been
discussed for more than 25 years (e.g., \opencite{BA2}) and are
now commonplace features of interplanetary exploration, as evidenced
by the Galileo and Cassini missions.

\begin{figure} 
\includegraphics{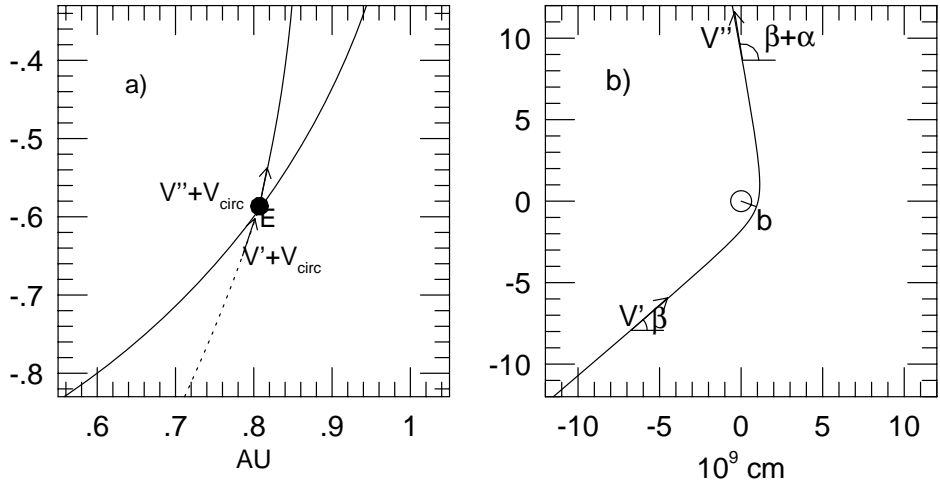}
\caption{
Geometry of orbital encounters and gravity assists for a
generic encounter of the {\bf O} and the Earth.  a) Encounter in the
heliocentric frame.  The distance
scale is in AU. b) Encounter in the planet-centered frame,
in which the object's flyby past the planet is a hyperbola with
close-approach distance $b$ and turning angle $\alpha$.  The distance scale is
in units of 10$^9$ cm, for a typical Earth encounter.
}
\end{figure}

\begin{figure} 
\includegraphics{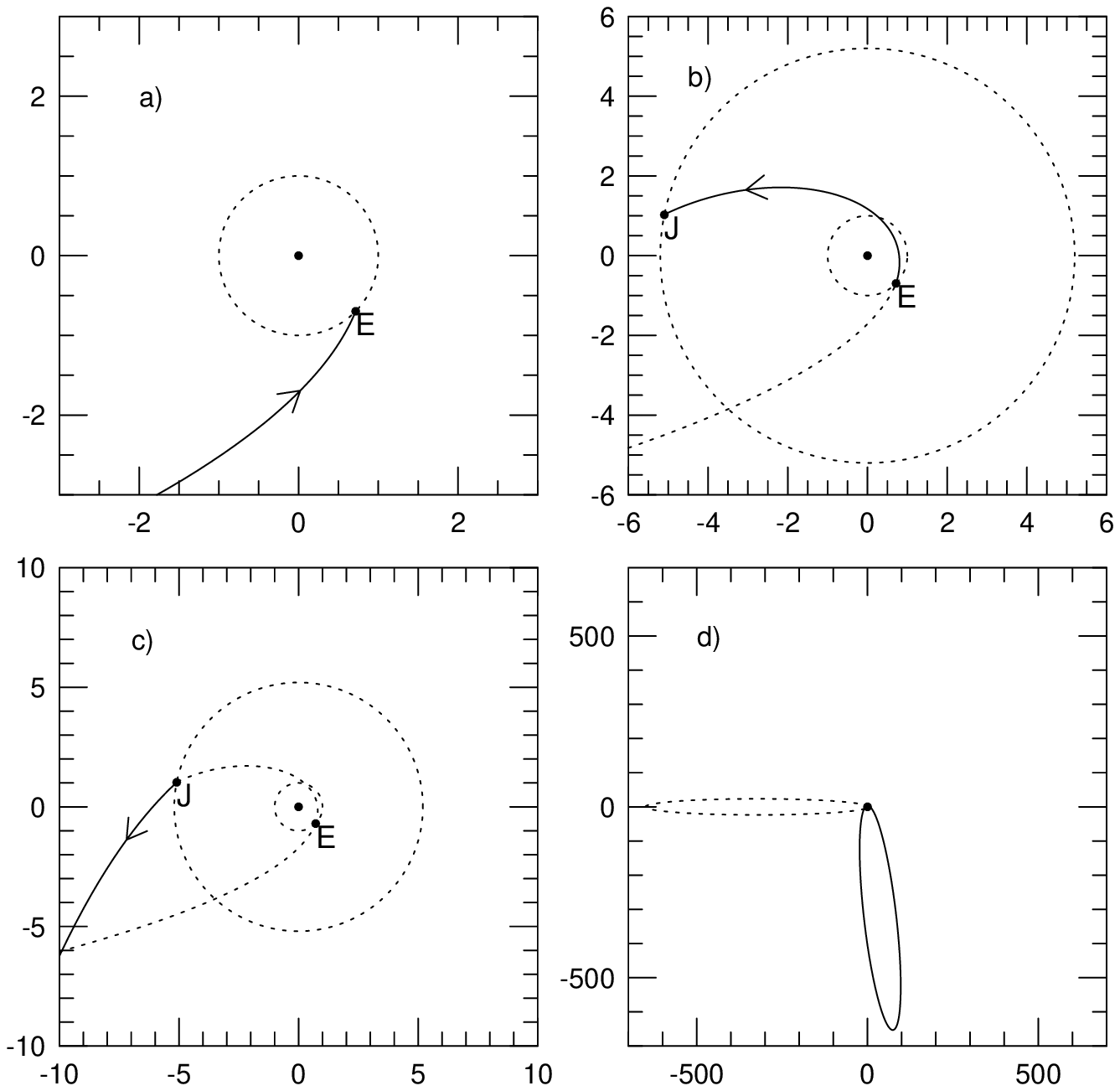}
\caption{
Layout of successive encounters for the Earth-Jupiter scheme,
for an orbit with initial aphelion $R_0$ at 650 AU and aphelion tangential
velocity $V_0 = 6000$ cm s$^{-1}$.  Note the
changes of scale from frame to frame. a) Initial orbit to Earth encounter
b) Orbit post-Earth to Jupiter c) Orbit post-Jupiter d) Initial (dotted) and
return (solid) orbits compared.
}
\end{figure}

The underlying dynamics of the scheme are shown in Figs. 1 and 2. 
The object ``{\bf O}'', a suitable Kuiper belt object or main belt
asteroid, first encounters the Earth during an inward pass on its
initial highly elliptical orbit of large ${\cal O}(300$ AU)
semimajor axis. The encounter transfers energy from {\bf O} to the
Earth in standard gravity-assist fashion by passing close by the
leading limb of the Earth.  The resulting orbit of {\bf O} then
crosses the orbit of Jupiter; with proper timing, it will encounter
Jupiter on its outbound swing, pass by Jupiter, and regain the energy
it lost to the Earth.  It would appear, however, that {\bf O} also
gains angular momentum relative to its incoming orbit, so that the
return orbit is less elliptical than the initial one.  As a result, a
modest amount of energy must be expended to restore {\bf O}'s orbit to
its initial parameters, unless further encounters are included that can
minimize the required expenditure. 

As discussed below, larger orbits for
{\bf O} entail lower first-order energy requirements. On the other hand,
an orbit that is too large will have too long a period to work well in
the scheme (unless multiple objects are used).
For purposes of discussion, we use an orbit with a semi-major axis
of 325 AU, whose period of 5859 years is compatible with the $\sim$ 6000
year average period between encounters.  In fact, the Sun's brightening
is slow at first and then speeds up; thus the optimum orbit size will 
decrease with time.  

\subsection{Formulation} 

We begin by assuming that Earth and Jupiter are on circular orbits of
zero inclination with radii of 1 and 5.2 AU, respectively.  Obviously,
a more detailed study would have to take into account the actual
elements of the planetary orbits.  To outline the scheme, however,
the idealized case is adequate. The calculation
of energy transfer and orbital parameters is made using the so-called
``patched-conic'' approximation (\opencite{B}, \opencite{BA1}).
In this approximation, the orbit of the object is treated as a series
of two-body problems.  Far from planets, {\bf O} follows a Keplerian
ellipse about the Sun.  At planetary encounters, {\bf O}'s path is
given by a two-body scattering encounter. Numerical integrations
of the full four body system indicate that this approximation gives
perfectly adequate results.

The incoming orbit of {\bf O} (characterized by subscripts 0) is
parameterized for convenience by an aphelion distance $R_0$ and
tangential velocity $V_0$.  Then the incoming angular momentum and
energy are given by
\begin{equation}
c_0 = R_0V_0,\qquad h_0 = {V_0^2\over 2} - {\musun\over R_0},
\end{equation}
where $\musun= GM_{\odot}$. Assuming $h_0<0$, the semi-major axis,
eccentricity, and longitude of perihelion of the orbit are
\begin{equation}
a_0 = -{\musun\over 2h_0}, \qquad 
e_0 = \left(1+{2h_0c_0^2\over \musun ^2} \right)^{1/2} , 
\qquad {\rm and} \qquad \omega_0 = 0 \, . 
\end{equation}
For an Earth encounter, the eccentricity must be large enough so that
the perihelion distance $a_0(1-e_0) < R_{\oplus}$, the orbital radius of
the Earth. If this constraint is satisfied, the Earth encounter takes 
place at longitude $\phi_E$, where 
\begin{equation}
\cos \phi_E = {1\over e_0}\left({p_0\over R_{\oplus}} -1\right) 
\qquad {\rm and} \qquad p_0 ={c_0^2\over\musun} \, .
\end{equation}
The detailed geometry of an encounter is illustrated in Fig. 1.
For an ``incoming'' (pre-perihelion) encounter, $\pi < \phi_E < 2\pi$.
The Earth's orbital velocity is $V_{\oplus} = (\musun / R_{\oplus})^{1/2}$.
The object's speed $V_E$, and tangential and radial velocities
$V_{TE}, V_{RE}$ at the encounter follow from conservation of angular momentum
and energy:
\begin{equation}
V_E = \left[2\left(h_0+{\musun\over R_{\oplus}}\right)\right]^{1/2},\qquad
V_{TE} = {c_0/R_{\oplus}},\qquad V_{RE}^2 = V_E^2 - V_{TE}^2.
\end{equation}
In the Earth's frame of reference, the velocities are
$V_{TE}^{\prime}=V_{TE}-V_{\oplus}, V_{RE}^{\prime}  = V_{RE}$, and the
encounter speed is given by $(V_E^{\prime})^2 = (V_{TE}^{\prime})^2
+ (V_{RE}^{\prime})^2$. The velocity vector of the encounter in
the Earth's frame makes an angle  $\beta_E$ with respect to the
orbital velocity of the Earth, where
$\cos\beta_E= V_{TE}^{\prime}/V_E^{\prime}$.

In the two-body treatment, the effect of the encounter is to turn the
velocity vector (in the Earth frame) through an angle $\alpha_E$, where
$\alpha_E$ depends on the encounter velocity and the impact parameter $B_E$.
The encounter can be timed to produce a specified minimum distance
from the Earth, $b_E$.  The minimum distance, the impact parameter $B_E$, 
and the turning angle $\alpha_E$ are related by
\begin{equation}
B_E = b_E\left(1+{2\muearth\over b_E (V_E^{\prime})^2}\right)^{1/2} 
\qquad {\rm and} 
\qquad \alpha_E = 2\tan^{-1}\left({\muearth\over B_E(V_E^{\prime})^2}\right).
\end{equation}
For an encounter in which {\bf O} loses energy and Earth gains energy, 
$\alpha_E>0.$ The post-encounter tangential velocity, 
from which the energy transfer is found,  is then given by 
$V_{TE}^{\prime\prime}=V_E^{\prime}\cos(\beta_E+\alpha_E)$.
Similarly, the post-encounter radial velocity is
$V_{RE}^{\prime\prime}=V_E^{\prime}\sin(\beta_E+\alpha_E)$.

The change in energy per unit mass of the object, 
from pre-encounter to post-encounter, is then given by 
\begin{equation}
\Delta Q_E = (1/2)[({\bf V}_E\cdot {\bf V}_E)_{\rm post} -
({\bf V}_E\cdot {\bf V}_E)_{\rm pre}] =
V_{\oplus}(V_{TE}^{\prime\prime} - V_{TE}^{\prime}).
\end{equation}
The corresponding change in the Earth's orbital energy is thus
$-M_O\Delta Q_E$, where $M_O$ is the mass of {\bf O}.  
As mentioned above, $\Delta Q_E$ will be negative 
(and hence the Earth will gain energy) if {\bf O} passes ``in front''
of the Earth.  The amount of energy transfer depends not only on the
minimum approach distance but also on the encounter geometry, i.e., 
$\beta_E$, and the encounter speed $V_E^{\prime}$, which in turn
depend on the longitude $\phi_E$ of the encounter.  Generally speaking,
the most effective encounters occur for near but not quite ``grazing''
encounters for which $\phi_E\sim \pm 0.5$ rad, not far from {\bf O}'s 
perihelion.  If $b_E$ can be taken as small as 10$^9$ cm (about 1.6
Earth radii), encounter transfer energies of up $\Delta Q_E \sim
10^{12}$ erg gm$^{-1}$ can be achieved. This value is approximately 
60\% of the maximum $\Delta Q_E = V_{\oplus}V_{\rm circ}$, where
$V_{\rm circ}$ is the circular velocity in Earth orbit at a radius
$b_E$ from the Earth's center.

We denote post-Earth-encounter quantities by the subscript 1. For 
these post-Earth variables, we use Cartesian vectors in the orbital 
plane, for which ${\bf R}_{\oplus} = (X_{\oplus}, Y_{\oplus}) =
(R_{\oplus}\cos\phi_E, R_{\oplus}\sin\phi_E)$. Similar formulae 
obtain for the post-Earth velocity (in the solar frame) ${\bf V}_1$.
In particular,
\begin{equation}
V_{x1} = V_{RE}^{\prime\prime} \cos\phi_E -
(V_{TE}^{\prime\prime} + V_{\oplus})\sin\phi_E,\qquad
V_{y1} = V_{RE}^{\prime\prime} \sin\phi_E +
(V_{TE}^{\prime\prime} + V_{\oplus})\cos\phi_E.
\end{equation}
The angular momentum is then given by 
$c_1{\hat z} = {\bf R_{\oplus}} \times {\bf V_1}$,
and the Laplace vector ${\bf P_1} = -\musun {\bf R_{\oplus}} / R_{\oplus} +
c_1{\hat z} \times  {\bf V_1}$.  With these forms, we obtain the 
orbital elements of the object {\bf O} in its post-Earth orbit, i.e., 
\begin{equation}
h_1 = {V_1^2\over 2} - {\musun\over R_{\oplus}},
\qquad a_1 = -{\musun\over 2h_1} , \qquad  
e_1 = {({\bf P_1}\cdot{\bf P_1})^{1/2} \over
\musun},\qquad \omega_1 = \tan^{-1}(P_{y1}/P_{x1}).
\end{equation}

Examination of the post-Earth orbital elements shows that the new
orbit of {\bf O} crosses the orbit of Jupiter.  We may therefore
schedule an encounter with Jupiter to regain energy lost by 
{\bf O} to Earth.  As before, treating the orbit of Jupiter as a circle 
and the orbit of {\bf O} as an ellipse in the plane, we find that the 
longitude $\phi_J$ of the Jupiter encounter is given by 
\begin{equation}
\cos (\phi_J-\omega_1)\ = {1\over e_1}\left({p_1\over R_{\jupiter}} -1\right),
\qquad {\rm where} \qquad p_1 ={c_1^2\over\musun} \, .  
\end{equation} 
The encounter of {\bf O} with Jupiter implies similar considerations
(with respect to the change in orbital parameters) as the encounter
with Earth. The orbital elements give us the tangential encounter
velocity $V_{TJ}^{\prime}$ (in Jupiter's frame) and the encounter
speed $V_{J}^{\prime}$.  Our initial idea was to set the encounter
geometry so as to yield an energy gain $\Delta Q_J$ (by {\bf O}) 
equaling the amount lost at the Earth. As noted below, however, a
more efficient encounter (in terms of the final velocity
change of {\bf O}) is one that yields a post-Jupiter orbit with an
aphelion equal to the original value $R_0$.  In that case, it is
simplest to search numerically for the desired Jupiter encounter
distance $b_J$.  The encounter geometry gives us $\beta_J$; the impact
parameter and $\alpha_J$ then follow from $b_J$ as above.

\begin{figure} 
\includegraphics{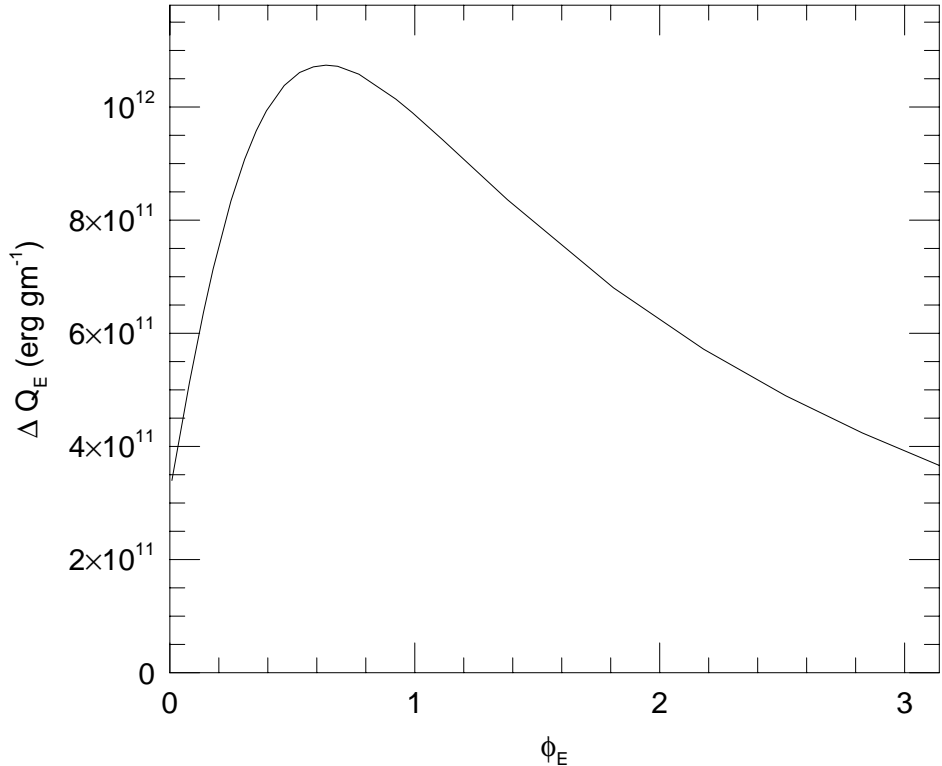}
\caption{
 Energy transfer (per unit mass of {\bf O}) $\Delta Q_E$ for encounters
as a function of $\phi_E$. All orbits have initial aphelia at $R_0$ at 650
AU.
}
\end{figure}

Finally, we must work out the orbital elements $a_R, e_R, \omega_R$
for {\bf O} on its post-Jupiter return orbit.  The succession of
encounters is shown in Fig. 2, for our example orbit with aphelion
at 650 AU and and aphelion tangential velocity $V_0 = 6000$ cm s$^{-1}$.
Figure 3 shows the energy transfer
for a collection of encounters ($b_E=10^9$ cm) as a function of $\phi_E$.

\begin{figure}
\includegraphics{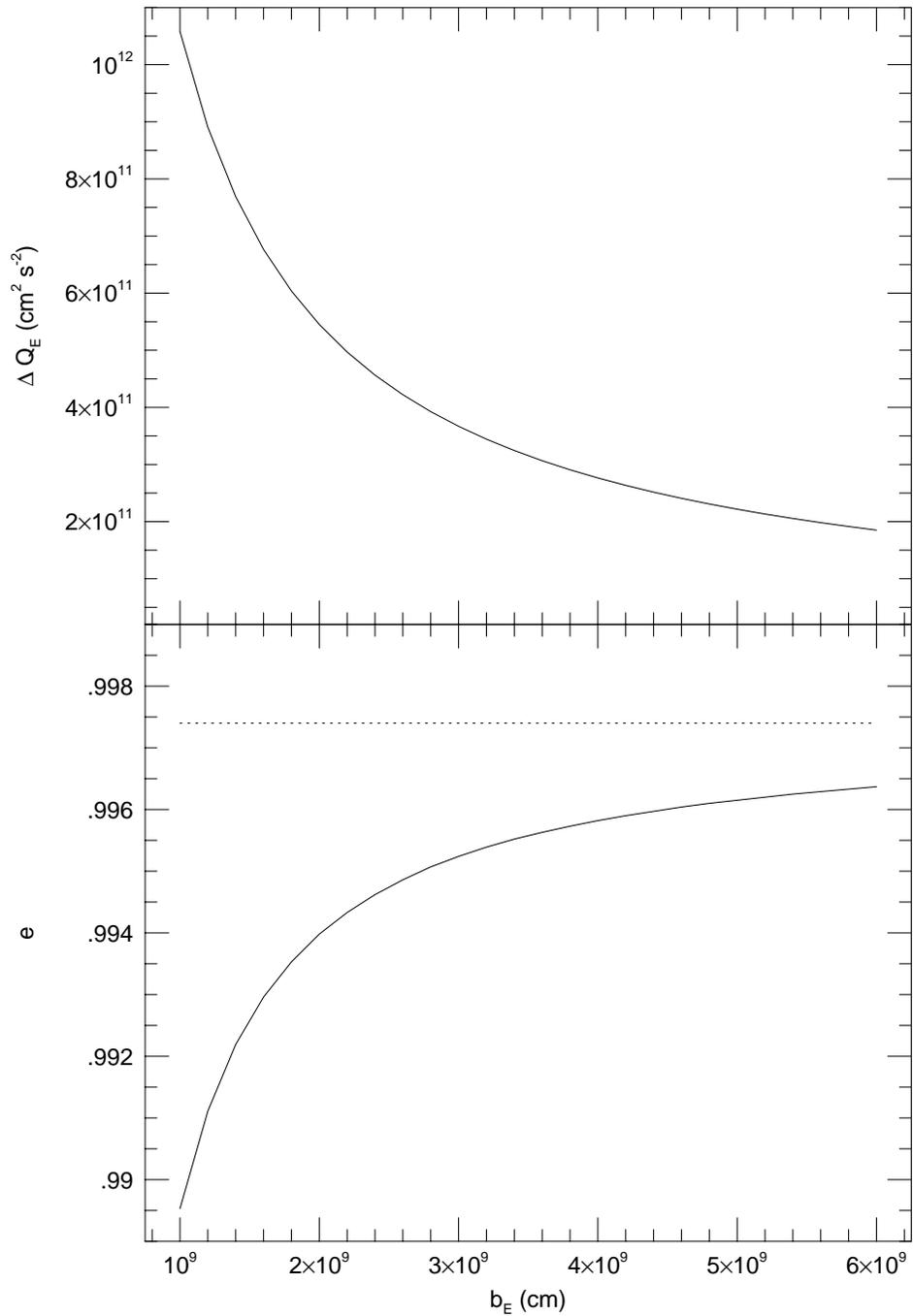}
\caption{
a) Energy transfer per unit mass of {\bf O} as a function of Earth
approach distance $b_E$ for orbits with $R_0 = 650$ AU and
$V_0 = 6000$ cm s$^{-1}$.  b) Eccentricities $e_0$
(dotted) and $e_R$ (solid) as a function of Earth approach distance $b_E$.
}
\end{figure}

In general, we find that $e_0 >  e_R$, and hence
the return orbit of {\bf O} has larger angular momentum than the
incoming orbit.  While we have not rigorously proven this result, it
seems intuitively understandable: the ``lever arm'' associated with
the Jupiter encounter (i.e., the radius of Jupiter's orbit) is 
$\sim 5$ times larger than that of the Earth encounter, resulting in a 
larger angular momentum. This finding spoils some of the neatness of
the scheme. It is possible to reduce the mismatch of angular momentum
by increasing the Earth encounter distance $b_E$, but this change
reduces the efficiency of the encounter, as $\Delta Q_E \propto
1/b_E$.  An example of this behavior is shown in Fig 4. If the
return orbit were identical to the incoming orbit (modulo its
orientation), a mechanism could be set up to recycle the object for 
an indefinite number of passes with a very low energy expenditure.   

The angular momentum of {\bf O} can be restored to its incoming value
by means of a ``course correction'' at aphelion.  Numerical
experimentation suggests that the most efficient scheme is to adjust
$b_J$ so as to produce a return orbit with the same aphelion as the
initial orbit: $a_R(1+e_R) = a_0(1+e_0)$.  The required velocity
change is then simply $\Delta V_R = c_R/R0 - V_0$, applied so as
reduce the tangential velocity of {\bf O} to its original value $V_0$.
This velocity correction is similar to the well-known Hohmann maneuver
used to transfer from one circular orbit to another with least velocity
change.  Since the change $\Delta V_R$ is inversely proportional to
the aphelion distance, it is advantageous (from the point of view of
least energy expenditure) to arrange for {\bf O}'s orbit to have the 
largest possible aphelion. On the other hand, the orbit must not be so large 
that its period is incompatible with the basic encounter timescale of
$\sim 6000$ years which is equivalent to a semimajor axis of $\sim 330$ AU.
For typical aphelia ${\cal O}(600)$ AU, the 
velocity change is $\Delta V_R\sim 6000$ cm s$^{-1}$.

\subsection{Multiplanet encounters post-Earth}

The considerations discussed above suggest that we consider the
possibility of scheduling multiple planet encounters after passage by
the Earth. This added complication can help optimize the scheme by
reducing the primary energy expenditure at the return-orbit
aphelion. An encounter with Saturn, immediately after the Jupiter
encounter, is a natural candidate.  Calculating the post-Saturn
orbital parameters follows in the manner outlined above.  We can then
search the parameter space of encounter distances with Jupiter $b_J$
and Saturn $b_S$ to minimize the velocity change $\Delta V_R$.

\begin{figure}
\includegraphics{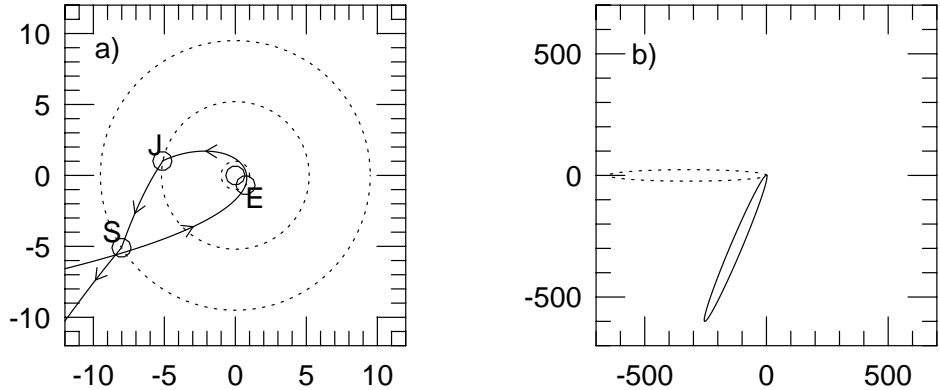}
\caption{
Successive encounters (cf. Fig. 2) for an Earth-Jupiter-Saturn
arrangement for an orbit with initial aphelion $R_0$ at 650 AU and aphelion
tangential velocity $V_0 = 6000$ cm s$^{-1}$.
a) Inner portion of the successive orbits. b) Initial (dotted) and
return (solid) orbits compared.
}
\end{figure}

\begin{figure}
\includegraphics{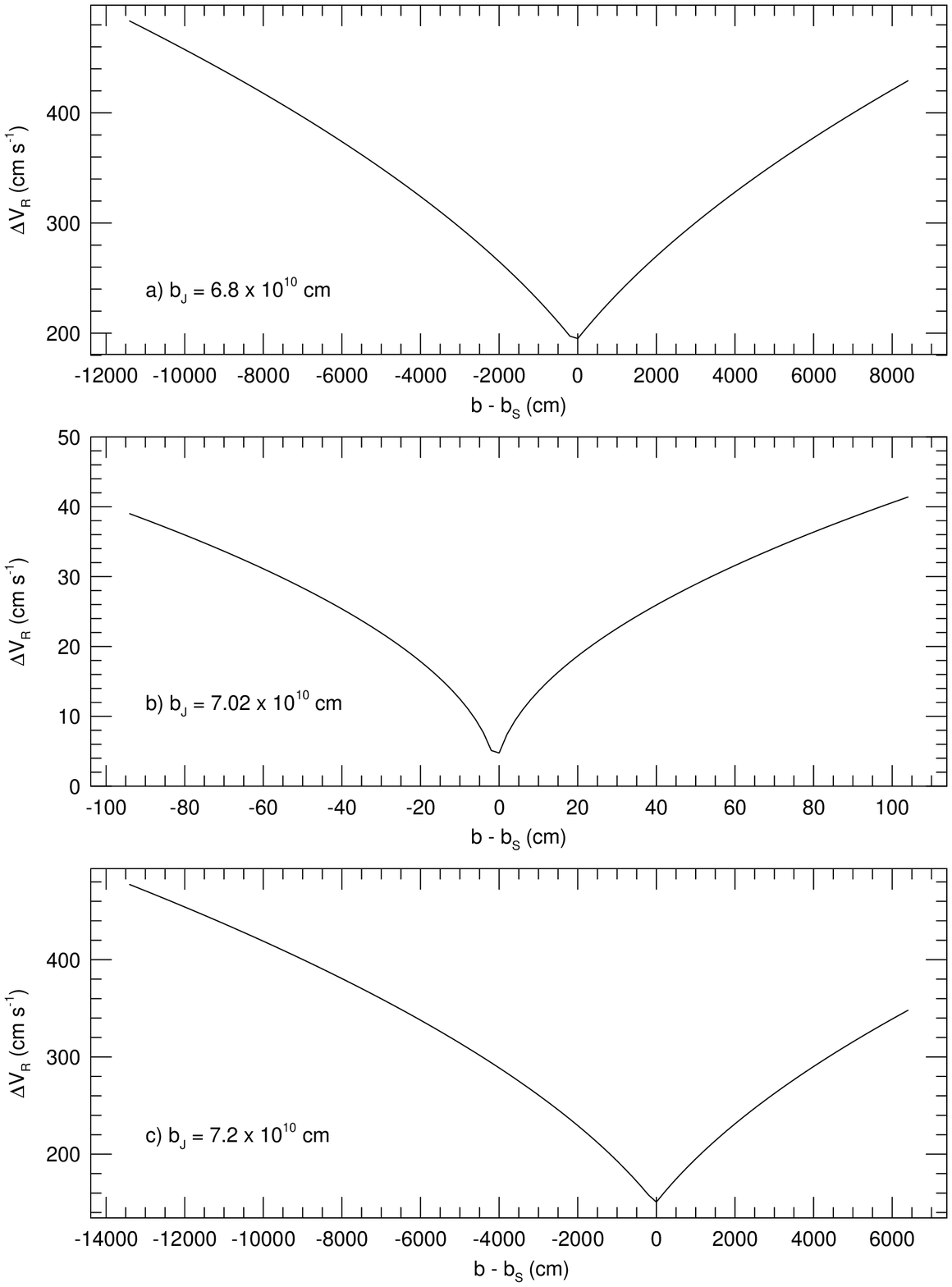}
\caption{
Velocity $\Delta V_R$ required to restore initial orbital energy and
angular momentum, as a function of difference $b-b_S$ from Saturn encounter
distance for three different values of Jupiter encounter distance $b_J$.
(Initial aphelion $R_0=$ 650 AU, aphelion tangential velocity $V_0 = 6000$
cm s$^{-1}$.)
a) $b_J = 6.8\times 10^{10}$, $b_S = 5.6106641326\times 10^{10}$ cm,
b) $b_J = 7.02\times 10^{10}$ cm, $b_S = 5.8751641993\times 10^{10}$ cm
c) $b_J = 7.2\times 10^{10}$ cm, $b_S = 6.1022703373\times 10^{10}$ cm
}
\end{figure}

We find that $\Delta V_R$ can in fact be reduced essentially to zero
by an arrangement where {\bf O} loses energy at the Saturn encounter; 
the distance $b_J$ must be decreased from its best single-encounter
value $\sim 1.73\times 10^{11}$ cm to compensate.  There is a dramatic
decrease of $\Delta V_R$ to nearly zero (${\cal O}$(10) cm s$^{-1}$)
for $b_J\sim 6.6\times 10^{10}$ cm. Figure 5 shows the orbital
geometry involved.  However, to find the minimum $\Delta V_R$ for any
specified value $b_J$ demands the specification of $b_S$ (or
vice-versa) to exceedingly high precision.  For example, reduction of
$\Delta V_R$ to $\sim 10$ cm s$^{-1}$ for $b_J = 6.6\times 10^{10}$
cm, requires $b_S$ to be specified to a precision of $\sim 10$ cm.
Less stringent specifications ${\cal O}(10^4)$ cm are sufficient if
reduction $\Delta V_R$ to a few meters per second is
satisfactory. Some examples are shown in Fig 6.

\begin{figure}
\includegraphics{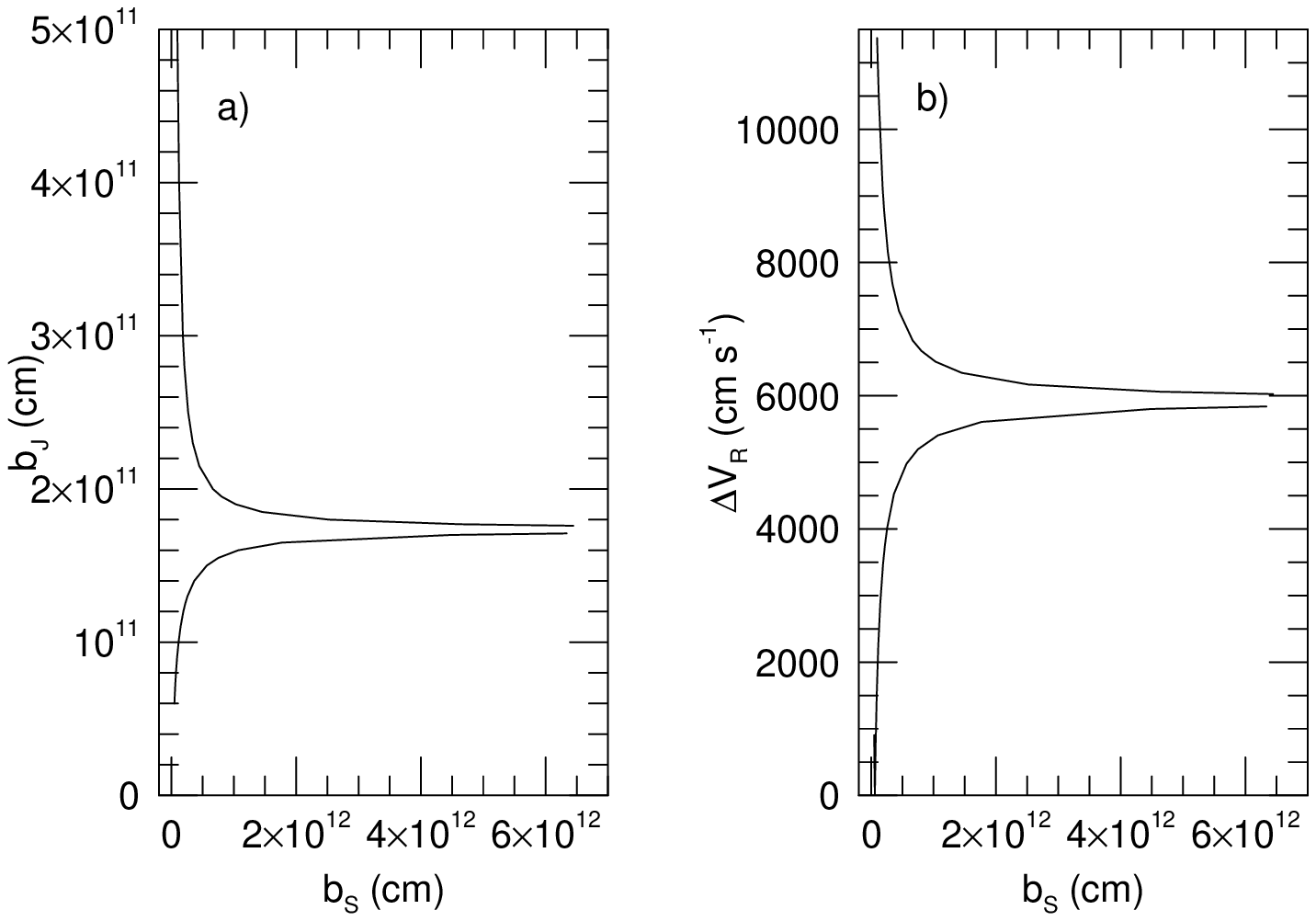}
\caption{
a) Jupiter encounter distance $b_J$ as a function of
Saturn encounter distance, required to minimize $\Delta V_R$.
(Initial aphelion $R_0=$ 650 AU, aphelion tangential velocity $V_0 = 6000$
cm s$^{-1}$.) The upper
curve is for encounters in which {\bf O} gains energy at Saturn; the lower
for encounters in which {\bf O} loses energy. b) Resulting minimum velocity
change $\Delta V_R$  for the encounters specified in a). Again, the upper
curve refers to energy-gaining encounters at Saturn and the lower for
energy-losing encounters.
}
\end{figure}

As a result, any realistic scheme will probably not attempt to
strictly enforce $\Delta V_R = 0$.  Nevertheless, it is interesting to
find that the ``first-order'' energy expenditure of the scheme can be
reduced in principle to a negligible amount. Figure 7 shows $b_J$ as a
function of $b_S$ for minimum $\Delta V_R$ transfers, and also the
resulting values of $\Delta V_R$.

\section{Other Issues} 
\label{sec:otheri} 

\subsection{Accuracy, Course Corrections, Energy Requirements}

In general, various additional effects will interfere with the
minimal-energy scheme outlined above.  These complications include
planetary orbital eccentricity, non-zero inclination angles, and
non-gravitational impulses. We will not attempt to develop a
sophisticated method of guidance for the object, although such methods
have been developed (to a very high degree of precision) for the space
program and planetary exploration (cf. \opencite{B}).  Instead, we will
merely make some estimates.

Planetary orbits have eccentricities and inclinations $e$ and $i$ 
(in radians) of a few times 10$^{-2}$. In order to accommodate these 
values, the  velocity changes will be $\sim 10^{-2}V$, or
about 10$^4$ cm s$^{-1}$ when applied to velocities of $\sim$ 10 km
s$^{-1}$, which would be typical of the region outside of Saturn's
orbit.  With sufficient planning, one can thus easily accommodate 
the departures of planetary orbits from circles in the same plane, 
as discussed so far. 

Undoubtedly, the need for other course corrections will occur.  High
accuracy is demanded at all critical stages. Closing velocities of 40
km s$^{-1}$ and encounter distances of 10$^9$ cm translate to accuracy
of ${\cal O}$(10-100) s in time of arrival at the Earth. We can get
some feeling for the size of velocity changes that are required in a
fairly simple way by making use of algorithms for the solution of
``Lambert's problem'' (\opencite{B}, \opencite{BA1}), which
consists of finding the velocity vector ${\bf v_0}$ needed to
produce a 2-body orbit that takes a mass from given position ${\bf
r_0}$ to ${\bf r}$ in a given time interval $\Delta t$.

We choose a target energy budget for velocity corrections of 
$\Delta V \sim 10^4$ cm s$^{-1}$.  We then use the algorithm for
Lambert's problem to compute differential velocity corrections along
the incoming orbit (before the Earth encounter) that yield changes in
arrival times $\Delta t$.  We find that  a $\Delta t$  of
${\cal O}$(10-100) s can be
accommodated fairly easily up to $\sim 10^5$ s before encounter.
Alternatively, much larger changes in arrival time can be allowed for
at greater distances; for example, at $\sim 0.5$ AU ($\sim 4\times
10^6$ s before encounter), the suggested budget could shift the
encounter time by $\sim 10^4$ s.  Of course, a more realistic
mission profile would not use up all the allowed energy for one
correction, and a target velocity change of $\Delta V \sim 10^3$ cm
s$^{-1}$ per course correction might represent a reasonable aim.

For the case of total $\Delta V=10^4$ cm s$^{-1}$, and for a 10$^{22}$
gm object, about 10$^{30}$ erg are required to enforce the velocity
changes for each encounter.  About 10$^{14}$ erg gm$^{-1}$ is
available from H$_2$O by deuterium-tritium fusion, assuming a
terrestrial D/H ratio \cite{PS}. Thus, $\sim 10^{16}$ gm
of ice must be proccessed for each encounter (at a minimum); this ice would
subtend a volume about 2.2 km on a side. For every encounter, this ice
volume represents about $10^{-6}$ of the mass of object {\bf O}; as a
result, the $\sim 10^{6}$ encounters necessary would consume the
deuterium of ${\cal O}(1)$ large Kuiper Belt object, assuming pure H$_2$O 
ice composition.  In addition, about twenty times as much rock mass is
needed to provide lithium for the production of tritium.   An object
of predominantly chondritic composition may thus be required if 
{\it in situ} production  is desired.
Obviously, much less processing is needed if {\it p-p}
fusion were available; in that case, a single object with
an associated processing and powerplant could be easily be used for the entire
project.

Non-gravitational forces are also an issue, and they potentially
demand energy expenditures above and beyond those that we have
discussed thus far.  An icy object, although attractive from the
standpoint of containing fusionable material and favorable initial
location in the outer solar system, will be more subject to this
problem than an stony or metallic asteroid.  On the other hand, a
large main-belt asteroid would have to be placed into a suitable orbit
starting from the inner solar system, where energy requirements are
high. 

\subsection{Timing}

The successful implementation of this scheme demands a reasonably
delicate interplay between orbital time scales of a thousand or more
years (for {\bf O}) and arrival times scheduled to the minute.  The
first major issue is how often one would expect to have Earth and
Jupiter (and Saturn) arrive in a particular configuration relative to
the argument of perihelion of {\bf O}. The object {\bf O} spends most
of its time ``hanging'' at aphelion. Small adjustments in trajectory
can thus be used to time the infall to correspond to the moment of
proper planetary alignment. With this flexibility, we could arrange
for  {\bf O} to arrive when Earth and Jupiter are in proper
position, an alignment that takes place every 13 months or so.  
As a result, two-planet encounters are easily realized; three-planet 
encounters (e.g., including Saturn) are a bit more difficult. 

In order for the most ideal version of our proposed mechanism to
operate, Earth, Jupiter, and Saturn must all be in the proper phases
of their orbits when {\bf O}  makes its passage through the inner solar
system. We can safely assume that minor corrections to the orbit of {\bf O} 
can delay its arrival in such a way that  {\bf O}'s orbital phase is
suitable for moving the Earth. The three planets, however, must have
the proper phase relative to each other. The conditions for this phase
alignment can be written in the form 
\be
(\omega_{\jupiter} - \omega_{\saturn}) t = 2 \pi n \, , 
\ee 
and 
\be
(\omega_{\oplus} - \omega_{\jupiter}) t = 2 \pi k \, , 
\ee
where $n$ and $k$ are integers and where the $\omega_j$ denote 
the orbital frequencies of the planets in obvious notation. 

We first consider the case in which the orbital frequencies, 
or equivalently the orbital periods $P_j = 2 \pi/\omega_j$, 
are constant. Expressed in years, the planetary periods are 
approximately $P_{\oplus}$ = 1, $P_{\jupiter}$ = 11.86, and 
$P_{\saturn}$ = 29.28. 

The condition for a perfect alignment can be written in the form 
\be
{k \over n} = {(P_{\jupiter} - 1) P_{\saturn} \over 
(P_{\saturn} - P_{\jupiter}) } \, 
\ee
where we have used the orbital periods rather than 
the orbital frequencies. Since the orbital periods are 
known, the right hand side is a known dimensionless number, 
which has a value of about 18.25.  We can write this expression 
in the form 
\be 
{k \over n} = 18.d_1 d_2 d_3 d_4 d_5 \dots \, , 
\ee
where the $d_j$ denotes digits of the number (which in general will be
irrational).  The general mathematical problem is thus to represent a
real number (the right hand side above) with a rational approximation. 
For a given specified accuracy (i.e., for a given number of decimal
places in the above expression), we need a minimum size of the
integers $k$ and $n$. The integer $k$ is roughly the number of Earth
orbits required to attain sufficient alignment, and hence is also the
approximately number of years between alignments (more precisely, $k$ 
measures time in units of $P_{\oplus} (1 - P_{\oplus}/P_{\jupiter})^{-1}$ 
$\approx$ 1.09 years). 

One possible choice for the alignment integers is thus $k$ = $18 d_1
d_2 d_3 d_4 d_5 \dots d_j$ and $n = 10^j$, where the last digit
represented is the $jth$ one. The time interval between alignments 
is thus about $\tau = P_{\saturn} P_{\jupiter} n /
(P_{\saturn} - P_{\jupiter})$ $\approx$ $20 \times 10^j$ years. 

The above argument shows that a solution exists.   
A compromise  must be made, however. In order to increase the accuracy 
of the alignment, we need longer time intervals between encounters. 
But we also need enough encounters per unit time to move the Earth 
before the Sun compromises the biosphere. 

Although the alignment condition will vary as Earth changes its
orbital parameters due to the asteroid encounters, the orbital period
of Earth only changes by a factor of two and hence the accuracy
requirements will be of the same order of magnitude for the entire
migration time interval.  The accuracy needed for alignments is
determined by the accuracy needed for the secondary encounters at
Jupiter and Saturn. We can assume that the orbit of {\bf O}
will be tuned to interact with Earth at just the right impact
parameter, but no course adjustments can be made before {\bf O}
reaches the outer planets. Jupiter and Saturn thus must be in the
right place to an accuracy of a few planetary radii (say $\ell \sim
10^{10}$ cm).  This constraint implies that we know the orbital phases
to a relative accuracy of about $\ell/r$ $\sim 10^{-4}$; we must take
$j=4$ at the very least, but we would like to use 
$j=5$ or even 6 in an ideal case. 
With $j=4$, for example, the time interval between encounters
(alignment opportunities) is about 200,000 years.  During the
allowed few billion year time period (until the biosphere is
compromised), we thus only get about 10,000 encounters. But, as 
discussed above, we need nearly a million encounters to successfully 
move the Earth to a viable larger orbit. 

We must thus view the problem the other way around: In order for
migration to occur within a few billion years, we must have encounters
every few thousand years.  For this frequency of encounters, the
largest allowed value of the integer $k$ is about 1000 (for example,
one obvious approximation would be $k$=1825 and $n$=100).  With this
level of precision, we can tune the encounter so that Earth and
Jupiter are in the right place, but Saturn will generally be in the
wrong phase of its orbit by an amount corresponding to a fraction
$\sim$0.02 of its orbit. The spatial displacement will be 0.02 $\times
2 \pi r$ $\approx$ 1.2 AU $\approx$ 3000 $R_{\saturn}$ (Saturn planetary
radii).  In this case, we can use Saturn to make course
corrections to {\bf O}'s orbit , but we cannot obtain perfect post-encounter
orbital elements (where the object has exactly the same energy and
angular momentum it started with).  

Therefore a  more realistic goal is to aim
at reducing the aphelion $\Delta V$ of {\bf O} to some small
value, or at least one that does not dominate the "energy budget".  
There is a range of possible Jupiter encounter parameters $b_J$
that yield a final  $\Delta V < 1000$ cm s$^{-1}$; these encounters
correspond to a range of $\sim$ 0.05 radian of encounters along 
Saturn's orbit, thus easing the timing requirement to a manageable level.

In addition  there are other considerations that mitigate the problem:
\begin{itemize}
\item Uranus and Neptune are available, giving three times as many 
opportunities as using Saturn alone.

\item Multiple objects can be used for energy transfer, though this
will probably raise the energy
requirements in proportion to the number of bodies involved.

\item Encounters need not be scheduled at the first opportunity
(as shown in Figs. 2 and 5). They can also be timed to occur after
multiple orbit passes at either intersection point of the orbits. The
object {\bf O} can be stored in temporary Chiron-like orbits as well.

\end{itemize}

\section{Discussion} \label{sec:conclude}

In this paper, we have investigated the feasibility of gradually
moving the Earth to a larger orbital radius in order to escape from
the increasing radiative flux from the Sun. Our initial analysis shows
that the general problem of long-term planetary engineering is almost
alarmingly feasible using technologies that are currently under
serious discussion.  The eventual implementation of such a program,
which is moderately beyond current technical capabilities, would
profoundly extend the time over which our biosphere remains viable.

The main result of this study is a theoretical description of a
workable scheme for achieving planetary migration. This scheme is
applied to the particular case of the Earth.  Solar system bodies,
such as large asteroids or Kuiper Belt objects, can be used to
move Earth over the next billion years.  These secondary bodies are
employed in a gravity-assist mechanism to increase the Earth's orbital
energy and thereby increase its distance from the Sun. The most
favorable orbits for the secondary bodies have a large semi-major
axis, typically hundreds of AU; with this relatively high ``leverage
factor'', the large requisite energy transfer can be achieved.

An important aspect of this scheme is that a single Kuiper
Belt object or asteroid
can be employed for successive encounters. In order to move the Earth
at the required rate, approximately one encounter every 6000 years (on
average) is needed (using objects with mass $\sim 10^{22}$ gm).  Due
to the acceleration of the Sun's luminosity increase, the encounters
must be more frequent as the Sun approaches the end of its
main-sequence life.  In order to use the same secondary body for many
encounters, modest adjustments in its orbit are necessary. However, by
scheduling the secondary body to encounter additional planets (e.g.,
Jupiter and/or Saturn) in addition to the primary Earth encounter, the
energy requirements for orbital adjustment at the object's aphelion
can be substantially reduced. In particular, the energy consumed by
such course corrections is not likely to dominate the energy budget.

Any serious proposal for planetary engineering, or any large-scale
alteration of the solar system, raises important questions of
responsibility (see \opencite{PS}).  Compared with other
astronomical engineering projects, this scheme has both positive and
negative aspects.  For example, although no massive alteration of
planetary environments is proposed, this scheme would consume a number
of large Kuiper Belt objects.

A great deal of energy must be expended to implement this migration
scheme. However, the energy needed to move Earth is relatively
modest compared to that needed for interstellar travel. For example,
an optimistic minimum energy expenditure is about 10$^{36}$ erg, which
corresponds to the kinetic energy of a $\sim 10^{23}$ gm object moving
at a velocity of 50 km s$^{-1}$ (this mass is less than 10$^{-4}
~M_{\oplus}$).  As a means of preserving the entire biosphere, this
scheme is thus highly efficient compared to interstellar migration,
even if we have underestimated the energy requirements by many orders
of magnitude. The energy requirements and overall ease of
implementation also compare favorably with various terraforming
projects \cite{PS}.

As noted near the beginning of this paper, the required change in orbital 
energy of the Earth is $\sim 9\times 10^{39}$ erg.  In the basic scheme we
have outlined, the energy is essentially transferred from Jupiter to the
Earth.  As a result, Jupiter's semi-major axis $a_{\jupiter}$ decreases by 
$\Delta a = a_{\jupiter}\Delta E/E_{\jupiter} 
\sim 5.3\times 10^{-3}a_{\jupiter}$, where 
$E_{\jupiter}=GM_{\odot}M_{\jupiter}/2a_{\jupiter}=1.7\times 10^{42} $ erg 
is Jupiter's orbital energy; this change amounts to $\sim 0.03$ AU.  
While small, this orbital change could destabilize some asteroidal or other 
orbits by the shift of position of Jupiter's orbital resonances.  The 
multi-planet scheme would involve similar-sized orbital changes for Jupiter 
and Saturn (or other planets).

Potentially more serious questions involve the rotation rate of the
Earth and the Moon's orbit.  We expect that {\bf O} will raise a tide
in the Earth during its encounter. The tide could be substantial; although
{\bf O} would be a relatively small  body, the closeness of its passage
means that the transient forcing potential would be ${\cal O}(10)\times$ as
strong as that of Moon, for a $10^{22}$ gm body passing $10^9$ cm from
the Earth's center.  Calculating the size and phase of the 
tide would require detailed work, but qualitatively we would expect
any tidal bulge to lag in phase behind {\bf O}, as {\bf O} moves
more quickly than the Earth rotates.  This in turn implies a spin-up
of the Earth (similar reasoning accounts for the the spin-down of the
Earth by the Moon).  Given the very large number of encounters planned,
a serious increase in the Earth's rotation rate could result.  

However, the above picture, leading to spin-up, takes place only for
``incoming'' encounters, such as depicted in Figs. 1 and 2.  The symmetry
of the encounters equally allows ``outgoing'' encounters, in which {\bf O} 
passes by the Earth after its perihelion.  Such encounters also pass by the 
Earth's leading limb from inside the Earth's orbit. They are thus retrograde 
with respect to the Earth's rotation, and the same considerations as above now 
lead to spin-down of the Earth rather than spin-up. Thus, by careful planning 
of encounters, we can cancel any unbalanced torques exerted on the Earth.

As for the Moon, reasoning by analogy with cases of stellar binaries
and third-body encounters suggests that the Moon will tend to become unbound
by encounters in which {\bf O} passes inside the Moon's orbit. 
(As well, there is the non-zero probability of collisions between 
{\bf O} and the Moon, which must be avoided.) Again,
detailed quantitative work needs to be done, but it seems that the
Moon will be lost from Earth orbit during this process.  On the other hand,
a subset of encounters could be targeted to ``herd'' the Moon along with
the Earth should that prove necessary.  It has been suggested 
(cf. \opencite{WB}) that the presence of the Moon maintains the
Earth's obliquity in a relatively narrow band about its present value and
is thus necessary to preserve the Earth's habitability.  Given that the
Moon's mass is 1/81 that of the Earth, a similarly small increment of
the number of encounters should be sufficient to keep it in the Earth's
environment.

The fate of Mars in this scenario remains unresolved. By the time this
migration question becomes urgent, Mars (and perhaps other bodies in
the solar system) may have been altered for habitability, or at least
become valuable as natural resources.  
Certainly, the dynamical consequences of significantly re-arranging
the Solar System must be evaluated. For example, recent work
by \inlinecite{IMW} has shown that if the Earth were removed
from the Solar System, then Venus and Mercury would be destabilized
within a relatively short time. In addition, the Earth will traverse
various secular and mean-motion resonances with the other planets as
it moves gradually outward. A larger flux of encounters might be
needed to escort the Earth rapidly through these potential trouble
spots.  In this case, additional solar
system objects may require their own migration schemes.  

This technology could also be used, in principle, to move other planets
and/or moons into more favorable locations within the solar system,
perhaps even into habitable zones. As an application, the basic
mechanics of this migration scheme could be employed to clear
hazardous asteroids from near-Earth space.  There is also the possibility
of using Kuiper-belt objects as resources themselves (e.g. of
volatile materials); gravitational-assist schemes could perhaps deliver
materials to useful locations with a minimum expenditure of energy.

An obvious drawback to this proposed scheme is that it is extremely
risky and hence sufficient safeguards must be implemented. The
collision of a 100-km diameter object with the Earth at cosmic
velocity would sterilize the biosphere most effectively, at least
to the level of bacteria.  This danger cannot be overemphasized.

\begin{acknowledgements}

DGK and GL are supported by a NASA astrophysics theory program which
supports a joint Center for Star Formation Studies at NASA-Ames
Research Center, UC Berkeley and UC Santa Cruz.  FCA is supported by
NASA Grant No.~NAG~5-2869 and by funds from the Physics Department at
the University of Michigan.

\end{acknowledgements}

\pagebreak
\end{article}
\end{document}